\documentclass[twocolumn]{aa} 

\usepackage{txfonts}
\usepackage[utf8]{inputenc}
\usepackage{dirtytalk}
\usepackage{amsmath}
\usepackage{amsfonts}
\usepackage{amssymb}
\usepackage{graphicx}
\usepackage{dcolumn}
\usepackage{subcaption}
\usepackage{bm}
\usepackage{latexsym}
\usepackage{epsfig}
\usepackage[normalem]{ulem}
\usepackage[draft,colorlinks=true, citecolor=blue]{hyperref}
\usepackage[dvipsnames]{xcolor}
\usepackage{todonotes}
\hypersetup{colorlinks=true,allcolors=blue}

\def\aap{A\&A}

\def\apjs{ApJS}
\def\apjl{ApJL}

\renewcommand{\vec}[1]{\bm{#1}}

\newcommand{\borg}{{\sc borg}}

\newcommand{\Msun}{\mathrm{M}_\odot}
\newcommand{\kms}{km~s$^{-1}$}
\newcommand{\kmsMpc}{km~s$^{-1}$~Mpc$^{-1}$}
\newcommand{\Mpch}{{$h^{-1}$~Mpc}}

   \titlerunning{Velocity correction for Hubble constant measurements from standard sirens}
   \authorrunning{Mukherjee et al. (2019)}

\begin{document} 
\definecolor{mygreen}{rgb}{0.0, 0.26, 0.15}
\newcommand{\bdw}[1]{\textcolor{mygreen}{\textbf{BDW: #1}}}
\title{Velocity correction for Hubble constant measurements from standard sirens} 

   \author{Suvodip Mukherjee
          \inst{1,2,3}\thanks{s.mukherjee@uva.nl, mukherje@iap.fr}
          \and
          Guilhem Lavaux\inst{1}
          \and
          François R. Bouchet\inst{1}
          \and
          Jens Jasche\inst{4}
          \and
          Benjamin D. Wandelt\inst{1,2,5}
          \and
          Samaya Nissanke\inst{3}
          \and
          Florent Leclercq\inst{6}
          \and
          Kenta Hotokezaka\inst{7, 8}
          }

   \institute{Institut d'Astrophysique de Paris (IAP), CNRS \& Sorbonne Université, UMR 7095, 98 bis bd Arago, 75014 Paris, France
            \and
             Institut Lagrange de Paris (ILP), Sorbonne Universités, 98 bis Boulevard Arago, 75014 Paris, France
             \and
             Gravitation Astroparticle Physics Amsterdam (GRAPPA),
Anton Pannoekoek Institute for Astronomy and Institute for High-Energy Physics,
University of Amsterdam, Science Park 904, 1090 GL Amsterdam, The Netherlands
		\and
             The Oskar Klein Centre, Department of Physics, Stockholm University, AlbaNova University Centre, SE 106 91 Stockholm, Sweden
             \and
             Center for Computational Astrophysics, Flatiron Institute, 162 5th Avenue, New York, NY 10010, USA
		\and
		Imperial Centre for Inference and Cosmology (ICIC) \& Astrophysics Group, Imperial College London, Blackett Laboratory, Prince Consort Road, London SW7 2AZ, United Kingdom
		\and
		Department of Astrophysical Sciences, Princeton University, Princeton, NJ, 08540, USA
		\and
		Research Center for the Early Universe, Graduate School of Science, University of Tokyo, Bunkyo-ku, Tokyo 113-0033, Japan
             }

       \date{\centering\today }

  \abstract{Gravitational wave (GW) sources are an excellent probe of the luminosity distance and offer a novel measure of the Hubble constant, $H_0$. This estimation of $H_0$ from standard sirens requires an accurate estimation of the cosmological redshift of the host galaxy of the GW source, after correcting for its peculiar velocity. Absence of an accurate peculiar velocity correction affects both the  precision and accuracy of the measurement of $H_0$, particularly for nearby sources. We propose  {a} framework to incorporate such a peculiar velocity correction for GW sources.  {A first} implementation of our method to the event GW170817 combined with the Very Large Baseline Interferometry (VLBI) observation leads to a revised value of $H_0= 68.3^{+ 4.6}_{-4.5}$ \kmsMpc{}.  While this revision is minor, it demonstrates that our method makes it possible for obtaining an unbiased and accurate measurements of $H_0$ at the precision required for the  standard siren cosmology.}
  

   \maketitle
%
\section{Introduction}
Attempts to make an accurate measurement of the expansion rate of the Universe at the present epoch, known as Hubble constant ($H_0$), is ongoing since the discovery of the expanding Universe \footnote{The International Astronomical Union has recently renamed the Hubble law as the Hubble–Lemaître law, in recognition of the pioneering contribution of Lemaître (\url{https://www.iau.org/news/pressreleases/detail/iau1812/}.)} by \citet{1927ASSB...47...49L,1931MNRAS..91..483L} and \citet{1929PNAS...15..168H}. Several complementary approaches measure its value with high precision \citep{Hinshaw:2012aka, Bennett:2012zja, Ade:2013zuv, Anderson:2013zyy, Ade:2015xua, Cuesta:2015mqa, Aghanim:2018eyx, Riess:2019cxk}. However,  current measurements of $H_0$ obtained using standard rulers anchored in the early Universe (Cosmic Microwave Background (CMB), Baryon Acoustic Oscillations (BAO)) \citep{Ade:2013zuv, Anderson:2013zyy, Aubourg:2014yra, Ade:2015xua, Macaulay:2018fxi} and Big Bang Nucleosynthesis 
(BBN) \citep{Addison:2017fdm, 2018MNRAS.480.3879A} differ from late-Universe probes using standard candles (supernovae (SN) type-Ia) \citep{2009ApJ...695..287R, Riess:2019cxk}, strong lensing from the H0LiCOW
project \citep{Wong:2019kwg} and using the angular diameter distance between the lensed images as a calibrator \citep{Jee:2019hah}.  A recent measurement of $H_0$ from the Carnegie-Chicago Hubble program by using the Tip of the Red Giant Branch (TRGB) as a calibrator for the SN type-Ia reduces the tension \citep{Freedman:2019jwv}. However, a more recent analysis by \citet{Yuan:2019npk} has claimed inaccuracies in the calibration of \citet{Freedman:2019jwv}, which again aggravates the tension. 

Taken at face value, this tension, statistically significant by more than $4$\,$\sigma$, would necessitate revision of the flat Lambda Cold Dark Matter ($\Lambda$CDM) model of cosmology \citep{Verde:2013wza, Bernal:2016gxb, PhysRevD.96.043503, Kreisch:2019yzn, Poulin:2018cxd, Lin:2019qug, Agrawal:2019lmo, Verde:2019ivm, Knox:2019rjx}. Whether this discrepancy is associated with systematic or calibration errors in either of the data sets or indicates new physics is currently a subject of intense debate. In this context, the spotlight has turned to standard sirens \citep{1986Natur.323..310S, Abbott:2017xzu} with binary neutron star mergers as an independent probe with the potential to reach the required percent-level precision to validate the low redshift ($z$) determination of $H_0$ \citep{Dalal:2006qt,2010ApJ...725..496N, Nissanke:2013fka,Chen:2017rfc, Feeney:2018mkj, 2018MNRAS.475.4133S, Mortlock:2018azx}. This potential depends crucially on whether the contamination from the peculiar velocity can be corrected at the required accuracy.  {The estimation of the cosmic velocity field was made in several studies \citep{1991MNRAS.252....1K,1992ApJ...391...16S,1994MNRAS.266..475H,1996ApJ...473...22D, 10.1046/j.1365-8711.1999.02514.x, 2000ASPC..201..223S, 2001MNRAS.326.1191B,10.1046/j.1365-8711.2001.04107.x,2004MNRAS.352...61H,10.1111/j.1365-2966.2004.08420.x,2005ApJ...635...11P, 2006MNRAS.373...45E, 10.1111/j.1365-2966.2011.19233.x,  10.1111/j.1365-2966.2011.18362.x,2012MNRAS.425.2880M,2012MNRAS.420..447T}}. In this paper, we propose a new framework to obtain an unbiased and accurate measurement of $H_0$ from GW observations.  {Our method relies on the \borg{} framework, for Bayesian Origin Reconstruction from Galaxies \citep{jasche_bayesian_2013,jasche2015,lavaux_unmasking_2016} to reconstruct the cosmic velocity field using the galaxy field observed in the redshift space. This framework is {quite} different from the methods used by \citet{Abbott:2017xzu}, which depends on the linear velocity estimates from \citet{Carrick:2015xza}. Along with the complete Bayesian posterior distribution of the velocity field available from the \borg{} framework, it is also useful in capturing the non-linear velocity component (as discussed in Sec. \ref{borg}), which goes beyond the framework by \citet{Carrick:2015xza}. Our method is an alternative way to incorporate the peculiar velocity corrections to the future GW sources in the sky patch of the 2M++ and SDSS-III samples as discussed in Sec. \ref{borg}. In order to make a reliable estimation of $H_0$ from  GW sources, it is essential to correct for peculiar velocity using multiple independent approaches to minimize the chance of any systematic bias. The proposal made in this work can be included along with other methods such as \citet{Carrick:2015xza}, and \citet{Springob:2014qja} for  future GW sources.} 

The paper is organised as follows, in Sec. \ref{basic} we discuss the low redshift probes to $H_0$ using standard candles and standard sirens. In Sec. \ref{luminosity-distance} and Sec. \ref{borg} we discuss the effect of peculiar velocity on  luminosity distance and redshift, and discuss a framework \borg{} to estimate it from the  cosmological observations.  
In Sec.~\ref{paves} we discuss the algorithm to incorporate peculiar velocity correction to the host of standard sirens. Finally in Sec. \ref{results} and Sec. \ref{conclusion} respectively, we obtain the revised value of $H_0$ from the event GW170817 \citep{Abbott:2018wiz} and discuss the applicability of our method to the future GW events.


\section{Low redshift probes of Hubble constant from standard sirens}
\label{basic}

All direct low $z$ measurements of $H_0$ depend on measuring the  luminosity distance  to the source, which is given by
 \begin{equation}\label{dl}
\begin{split}
d_L= \frac{c(1+z)}{H_0}\int_0^z \frac{dz}{E(z)},
\end{split}
\end{equation}
 in a homogeneous Friedmann-Lemaître-Robertson-Walker (FLRW) metric when assuming a stationary source and observer. Here $c$ is the speed of light, and $E(z)\equiv H(z)/H_0= \sqrt{\Omega_m (1+z)^3 + (1-\Omega_m)}$, within the framework of flat $\Lambda$CDM \citep{Hinshaw:2012aka, Bennett:2012zja, Ade:2013zuv, Anderson:2013zyy, Ade:2015xua, Cuesta:2015mqa, Aghanim:2018eyx} with  $\Omega_m= \rho_m/\rho_c$ (matter density $\rho_m$ today divided by the critical density $\rho_c= 3H_0^2/8\pi G$). At very low redshift $z$ the Hubble parameter $H(z)$ is nearly constant, this relation simplifies greatly and becomes independent of the background cosmological model
\begin{equation}\label{dl1}
  d_L= \frac{cz}{H_0}.
\end{equation}

  {Gravitational waves (GW) sources provide a new avenue to measure the luminosity distance since as they provide remarkable standard sirens which only assume that the general theory of relativity be a valid description; they do not require to be \lq standardized\rq\ using other astrophysical sources. Indeed, as pointed out by \citet{1986Natur.323..310S}, the distance to the GW sources can be measured without a degeneracy with its redshifted chirp mass $\mathcal{M}_z$, which is related to}  
{the physical chirp mass in the source frame by the relation $\mathcal{M}_z= (1+z)\mathcal{M}$. Source frame chirp mass is related to the mass of each of the compact objects, $m_1$ and $m_2$, by the relation $\mathcal{M}= (m_1m_2)^{3/5}/(m_1+m_2)^{1/5}$.} The GW strain in the frequency domain can be expressed for the two polarization states $h_+,\, h_\times$, by the relation \citep{1987thyg.book.....H, Cutler:1994ys,Poisson:1995ef,maggiore2008gravitational}
\begin{align}\label{gw-waveform}
\begin{split}
h_+ (f_z)&= \sqrt{\frac{5}{96}}\frac{G^{5/6}\mathcal{M}_z^2 (f_z\mathcal{M}_z)^{-7/6}}{c^{3/2}\pi^{2/3}d_L} \left(1+ \cos^2(i)\right) e^{i\phi_z},\\
h_\times (f_z)&= \sqrt{\frac{5}{96}}\frac{G^{5/6}\mathcal{M}_z^2 (f_z\mathcal{M}_z)^{-7/6}}{c^{3/2}\pi^{2/3}d_L} \cos(i) e^{i\phi_z+ i\pi/2},
\end{split}
\end{align}
where $G$ is the gravitational constant, $c$ is the speed of light, $f_z= f/(1+z)$ redshifted GW frequency, $\phi_z$ is the phase of the GW signal and $i\equiv \arccos(\hat L.\hat n)$ is the inclination angle of the GW source which is defined as the angle between the unit vector of angular momentum $\hat L$ and the line of sight of the source position $\hat n$. The above expression indicates the presence of a degeneracy between $d_L$ and $\cos(i)$ for  individual polarization states. However, the polarization states of GW can be measured as they have different dependency on $\cos(i)$ \citep{Holz:2005df,Dalal:2006qt, 2010ApJ...725..496N, Nissanke:2013fka}. The measurement of both polarization states of the GW signal requires multiple detectors, with different relative orientation between the arms of the detectors. 

Along with the luminosity distance measurement to the standard sirens, a measurement of the Hubble constant also requires an independent measurement of the redshift of the source. Binary neutron star, black hole--neutron star, supermassive binary black hole mergers are all expected to have electromagnetic counterparts. This can lead to the identification of the host galaxy of the GW source and the redshift to the GW source can be estimated from the electromagnetic spectra of the host using spectroscopic (or photometric) measurement by the relation $1+z= \lambda_o/\lambda_e$.\footnote{$\lambda_o$ and $\lambda_e$ are the measured and emitted wavelength of the light respectively.}  As a result, by using $d_L$ from the GW signal, and $z$ from the electromagnetic spectra, GW sources provide an excellent avenue to measure Hubble constant using Eq.~\eqref{dl1}.

\section{Luminosity distance and redshift in the presence of large scale structure} \label{luminosity-distance}

The luminosity distance to a source and its observed redshift in a homogeneous FLRW Universe is different from the one in presence of cosmic perturbations.\footnote{The perturbation in the homogeneous FLRW metric is caused by the presence of  structures in the Universe.} The presence of perturbations in the matter density leads to temporal and spatial fluctuations in the metric perturbations (through terms involving $\dot \Phi,\, \ddot \Phi$, $\vec{\nabla} \Phi$, $\vec{\nabla}^2 \Phi$), which are related to effects such as, but not limited to, the peculiar velocity of the source and of the observer, the gravitational redshift, the integrated Sachs-Wolfe and gravitational lensing  \citep{1987MNRAS.228..653S, Kolb:2004am, Barausse:2005nf, Bonvin:2005ps}. 

The observed redshift to the source also differs from the cosmological redshift due to the contributions from the difference in the peculiar velocity between the observer $\vec v_{o}$ and the source $\vec v_s$ and also to the gravitational redshift. At  low redshift, the most important contributions arise from the difference in the velocity of the observer and source. The observed redshift $z_{obs}$ can be written in terms of the peculiar velocity $v_p= (\vec v_{s}-\vec v_{o}).\hat u$~\footnote{The line of sight vector $\hat u$ is directed from the observer to the source, i.e., $\vec v.\hat u \geq 0$ implies that the source is moving away from the observer.} by the relation
$\left(1+z_\text{obs}\right)= \left(1+z\right)\left(1+ \frac{v_p}{c}\right)$. 
The corresponding modification in Eq.~\eqref{dl1} is 
 \begin{equation}\label{dl2}
d_L= \frac{c z+ v_p}{H_0}.
\end{equation}
This implies that the contribution from the peculiar velocity leads to a bias in the inferred value of $H_0$, if not accounted for. For multiple sources with uncorrelated velocities, the effect of ignoring the peculiar velocity component in the average is to produce excess variance in the measurement of $H_0$. If we assume the distribution of the peculiar velocity field to be Gaussian with a variance $\sigma_v^2$, then the corresponding excess variance in the $H_0$ measurement for a source at distance $d_L$ becomes $\sigma^2_v/d^2_L$. The current frameworks to obtain $H_0$ from the standard sirens use a Gaussian prior on the peculiar velocity \citep{Abbott:2017xzu, Chen:2017rfc, Feeney:2018mkj}. We propose an algorithm to correct for the peculiar velocity contribution to standard sirens.

The peculiar velocity of a host galaxy, $v_p= v_h+ v_\text{vir}$, can arise from two components namely (i) the motion of the halo due to the spatial gradient in the gravitational potential $v_h$, and (ii) the virial velocity component $v_\text{vir}$  of the host galaxy inside the halo. The non-rotational velocity component of $v_l$ can be obtained from  linear perturbation theory as 
\begin{equation}
\vec{v}_l(\vec{x},z)= - \frac{2}{3}\frac{g(z)}{aH(z)\,\Omega_\text{m}(z)}\, \vec{\nabla}_{\vec{x}} \Phi ,
\end{equation}
where $\Phi$ is the gravitational potential, $g$ is the growth rate related to the linear growth factor $D$ by the relation $g= \text{d}\ln D/\text{d}\ln a$ and $a=1/(1+z)$ is the scale factor. At small scales, non-linear effects can be quite important and the previous relation becomes inaccurate. The virial velocity component, $v_\text{vir}$, can be related to the  mass of the halo, $M_h$, and the distance to halo center, $r$, by the relation $v^2_\text{vir} \propto M_{h}/r$, which holds for a virialized system. For a system with a constant mass density, the halo mass and $r$ are related by $M_h \propto r^3$. As a result, the virial velocity $v_\text{vir}$ is only related to the mass of the halo by the relation $v^2_\text{vir}\propto M^{2/3}_h$. This relation indicates that a galaxy with a heavier halo has a larger velocity dispersion than if it resides in a smaller halo. We use the following form, fitted to simulations, to estimate the velocity dispersion of the non-linear velocity component \citep{2001MNRAS.322..901S}
 \begin{equation}\label{vvir}
\sigma_\text{vir}= 476\, g_v\, (\Delta_\text{nl}(z)E(z)^2)^{1/6} \ (M_h/ 10^{15}\, \Msun h^{-1})^{1/3},
\end{equation}
where $g_v=0.9$, $\Delta_\text{nl}(z)= 18\pi^2 +60x -32 x^2$, and $x= \Omega_\text{m} (1+z)^3/E^2(z)-1$.

\section{Estimation of the velocity field using \borg{}}\label{borg}

The velocity field that we use in this work is produced by the ``Bayesian Origin Reconstruction from Galaxies" (\borg{}) probabilistic model of large scale structure, as currently applied to the 2M++ compilation \citep{Lavaux2011_TMPP,Jasche2019_PM} and the SDSS-III/BOSS survey \footnote{SDSS-III/BOSS is the \href{http://www.sdss3.org/surveys/boss.php}{Baryon Oscillation Spectroscopic Survey}.} \citep{2011AJ....142...72E,Dawson2013,Lavaux2019_BORG}. The \borg{} algorithm aims at inferring a fully probabilistic and physically plausible model of the three-dimensional cosmic matter distribution from observed galaxies in cosmological surveys
\citep[see e.g.][]{jasche_bayesian_2013,jasche2015,lavaux_unmasking_2016}. To that effect, the method solves a large-scale Bayesian inverse problem by fitting a dynamical structure formation model to data and inferring the primordial initial conditions from which presently-observed structures formed.
The \borg{} forward modelling approach marginalizes automatically over unknown galaxy bias, and accounts for selection and evolutionary effects.
The variant of \borg{} that we use for the 2M++ catalogue models the galaxy to dark matter bias using a 3-parameter function motivated by the analysis of $N$-body simulations \citep{neyrinckHaloBiasFunction2014}.  {We note that this bias relation is a local non-linear relation for which conventional understanding of bias on large scale does not apply. It is more akin to a halo occupation distribution for high luminosity. The data specifically selects that behavior as can be seen in Section~5.1 of \citet{Jasche2019_PM}. The reader should approach with caution any naive numerical comparison of \borg{} bias parameters to linear bias values found in the literature}.
Instead of the linear perturbation result in the previous section, \borg{} uses a full particle-mesh $N$-body solver to evolve the initial conditions to a dark matter density distribution at $z\,\approx \,0$  \citep{Jasche2019_PM} through direct integration of the Hamiltonian dynamics equation. 

The \borg{} model provides a set of points in the parameter space (dimensionality $\simeq 256^3$, for initial conditions, and a few more for additional bias parameters) that provides a numerical approximation of the posterior distribution of those parameters given the 2M++ observed galaxy distribution.  Once initial and final positions of dark matter particles are known, the velocity field is estimated using the Simplex-in-Cell estimator \citep[SIC,][]{Hahn2014_DMSHEET,Leclercq2017a}.  {The SIC estimator relies on a phase-space interpolation technique, which provides an accurate estimate of velocity fields without the need for commonly used but arbitrary kernel smoothing procedures, such as cloud-in-cell (CIC). It provides a physical picture for velocity fields even in low-density regions, which are poorly-sampled by dark matter particles, and takes into account the multi-valued nature of the velocity field in shell-crossed regions by proper stream averaging.}

 {\subsection{Validation of the \borg{} reconstruction algorithm}}

 {In this section, we discuss several validation tests of the \borg{} velocity field inference. In Section~\ref{sec:linear_theory_performance}, we assess the validity on the sole basis of the statistics of the reconstructed samples. In Section~\ref{sec:n_body_test}, we run a \borg{} analysis on a set of mock tracers to check the unbiasedness and the typical reconstruction errors.}

\begin{figure}
     \centering
        \includegraphics[width=\hsize]{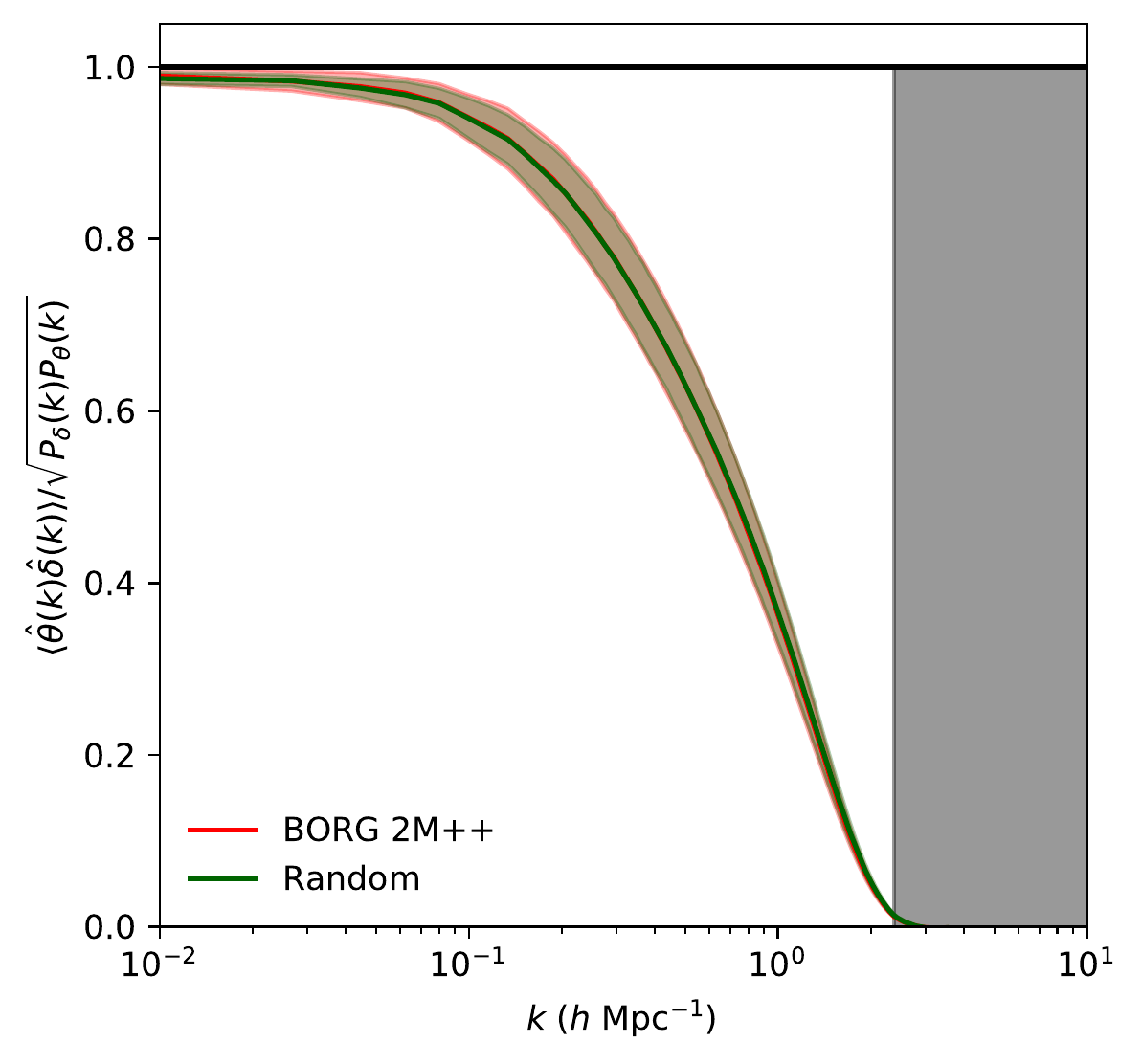}
 \caption{ {We show the correlation coefficient between the non-linear density and the negative  divergence of the non-linear velocity field, $\theta \equiv -\bm{\nabla}\cdot \bm{\textrm{v}}$. The red line is the ensemble average over the posterior distribution of the cross-correlation and the shaded dark red area is the one standard deviation according to the mean of that cross-correlation derived from the same distribution obtained by \borg{}. The green and shaded green region is the result of computing that same quantities on 80 random realizations. Apart from very minor differences, the two curves are on top of each other, which can be expected as the physical model is the same in both cases. The vertical grey band on the right side indicate the resolution limit of the \borg{} reconstruction. The correlation reach asymptotically one for the largest scales (small $k$ values). The red line is reaching a value of 99\% for the largest probed mode. As expected, this cross-correlation goes to zero at small scales (high $k$ value). \label{fig:borg_den_crosscor} }}
\end{figure}

\begin{figure}
    \centering
    \includegraphics[width=\hsize]{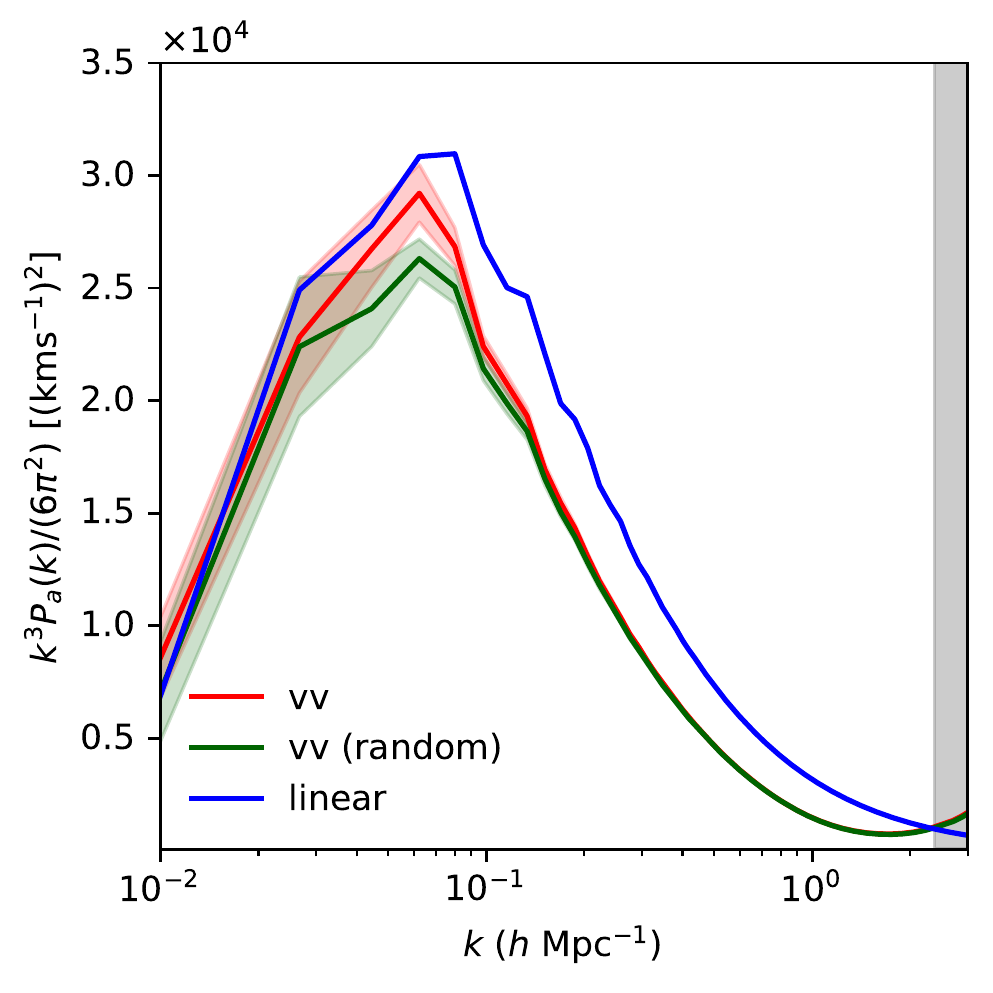}
    \caption{ {We show here the ensemble average power-spectrum of the non-linear velocity field scaled, alongside with one standard deviation contour in shaded area. The line and shaded region in red is computed from the ensemble posterior distribution built from the 2M++ data. The line and shaded region in green are measured from 80 random realizations of a $\Lambda$CDM universe of the same size. Those random realizations were computed using the same $N$-body solver, but starting from a random sample of a Gaussian distribution with power-spectrum provided by the $\Lambda$CDM model. We show in thick blue line the power spectrum assuming linear theory. We note that large scales are unaffected by non-linearities. Intermediate scales ($k \gtrsim 0.1 h$~Mpc$^{-1}$, corresponding to $\sim 60h^{-1}$~Mpc) get non-negligible contributions from non-linear dynamics, as highlighted by the area under the curve. The vertical grey band on the right side indicate the resolution limit of the \borg{} reconstruction ($\sim 2.6 h^{-1}$~Mpc).}}
    \label{fig:borg_vel_powerspectrum_log}
\end{figure}

\begin{figure*}
    \includegraphics[width=\hsize]{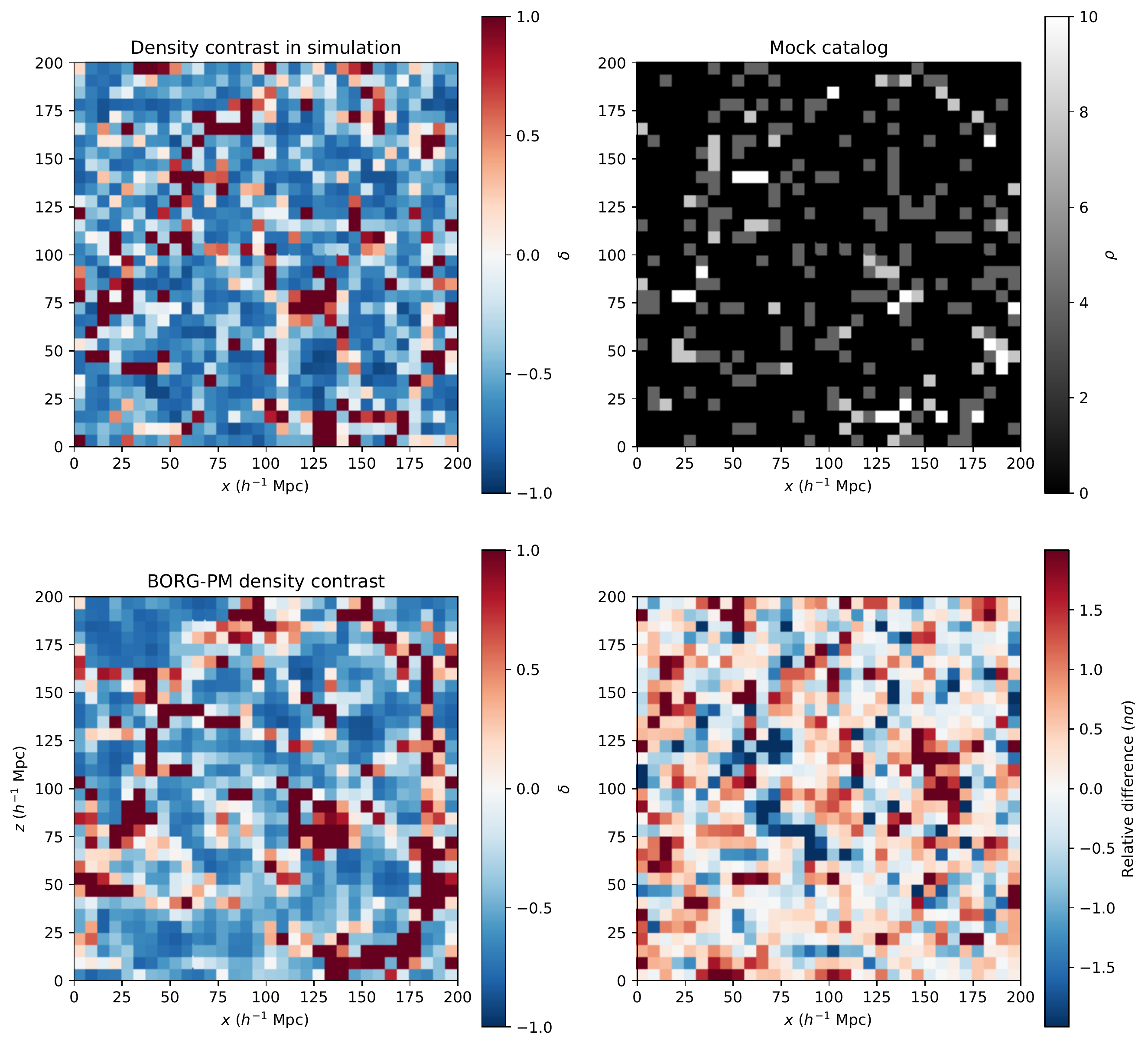}
    \caption{ {Tests of the \borg{} algorithm on a $N$-body simulation. {\it Top-left panel:} center slice of the density contrast computed at a resolution of $32^3$ from the $N$-body cosmological simulation used in our test. The simulation itself is obtained with $256^3$ dark matter particles with $\Lambda$CDM cosmology as used in \citet{Jasche2019_PM}. {\it Top-right panel:} Central slice of grid obtained by assigning objects of the mock catalog to a $32^3$ grid with a nearest grid point filter. {\it Bottom-left panel:}  Density slice computed from one posterior sample obtained by \borg{}. We note the clear spatial correlations between the density field of the fiducial simulation and the inferred sample, despite the very low sampling rate of the mock catalog. {\it Bottom-right panel:} Difference between the  \borg{} reconstruction and the simulation truth divided by the standard deviation per voxel estimated from the posterior distribution. The \borg{} posterior covers the truth and gives a conservative estimate of the uncertainty.  }}
    \label{fig:borg_pm_density_reconstruction}
\end{figure*}

\begin{figure}
    \includegraphics[width=\hsize]{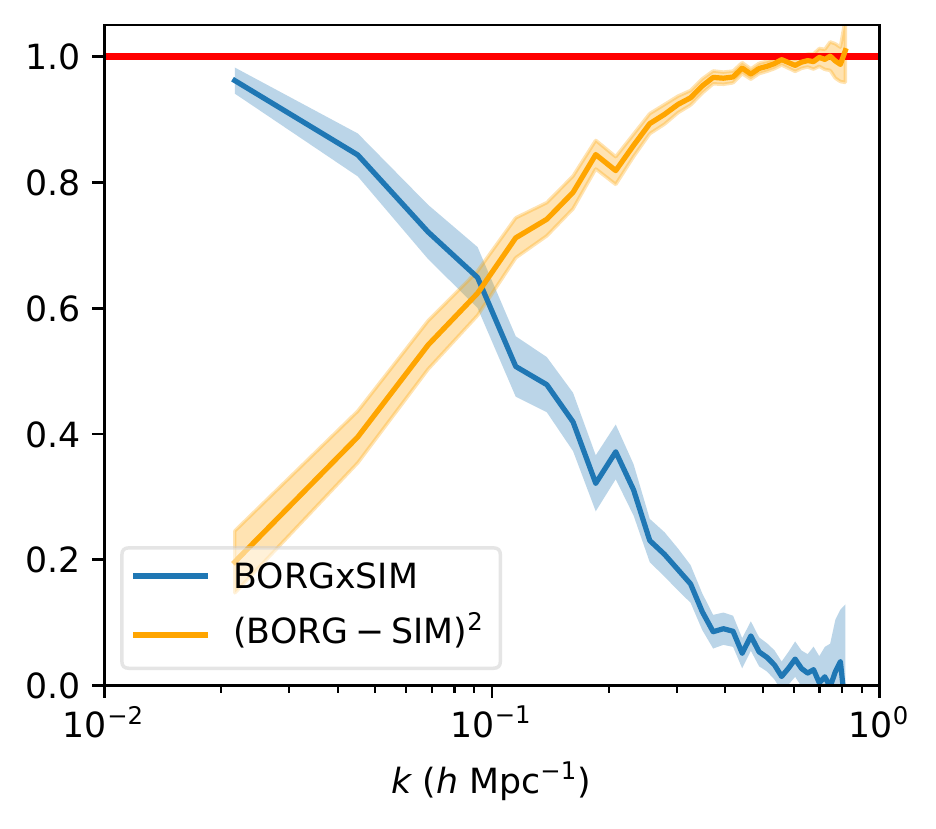}
    \caption{ { We show the correlation (blue color) and error residual (orange) for the density field inferred by \borg{} compared to the simulation density field. We have computed the mean and one standard deviation using the posterior distribution sampled by the \borg{} chain. Both curves are normalized by the power spectrum of each field.  We see that indeed when correlation is weak, we get the maximum relative variance.}}
    \label{fig:correlation_sim_borg_density}
\end{figure}

 {\subsubsection{Comparisons to performance of linear perturbation theory}\label{sec:linear_theory_performance}
}

 \citet{Carrick:2015xza} is commonly used for correcting the  {estimated cosmological} redshift of observed objects in the Nearby Universe. This model relies on computing the inverse Laplacian operator on a luminosity weighed distribution of the 2M++ galaxies. This model has been fitted to observed distances of galaxies derived from the SFI++ \citep{Springob:2007} and the First Amendment A1 SNIa data \citep{Turnbull:2012}. The \borg{} model goes beyond this approach. It is based on deriving a physically meaningful velocity field {, beyond linear perturbation theory,} based on several pieces of information, including the same 2M++ data-set and cosmology. 
 
 The performance and need of going beyond linear perturbations has been extensively discussed in the past. The differences between the density and velocity power spectra have already been computed analytically on large scales \citep{Scoccimarro:2004} and observed in simulations \citep{Chodorowski:2002,Scoccimarro:2004,Jennings:2011,Jennings:2012}, but they are challenging to derive from observational data. This  additionally motivates  the \borg{} approach. \citet[][figure 1]{Leclercq2015} has shown that \borg{}'s constrained simulations of the density field are statistically consistent with independent random realizations. We further emphasize this point here by analyzing the capability of going beyond the linear level for the velocity field. In that respect, we show  
the cross-correlation between the density and $\theta = - \bm{\nabla}\cdot\bm{\textrm{v}}$ in Figure~\ref{fig:borg_den_crosscor} for constrained realizations and for random simulations generated using the same particle mesh implementation and cosmological setup. 
There are no visible differences between the cross-correlations between the inferred 2M++ fields and the ones for random realizations. This provides evidence that the density de-correlates from the velocity divergence in the way that is expected from the non-linear gravitational clustering physics of dark matter. We also show in Figure~\ref{fig:borg_vel_powerspectrum_log} the contribution of the different scales to the total variance of the velocity field. We note that pure Eulerian linear perturbation theory (solid blue line) is systematically above the curve obtained from constrained 2M++ realizations (shades of red) and the random velocity field power spectrum (shades of green) for the range of scales of interest. That behaviour of course changes on smaller scales for which random motions become significant, as is hinted by the crossing of the two curves at $k\simeq 2h$~Mpc$^{-1}$.  {For the scales $k<0.8h$~Mpc$^{-1}$, the 2M++ constrained realization matches well with the pure Eulerian linear perturbation theory within the expected variance (denoted in the red shaded region). The drop of power at that scale for the mock and constrained realizations is related to the observation on non-linearities in the power spectrum as pointed out in \citet{Jennings:2011}.}

 {%
The two above points show the limits of using the density field as a predictor for the velocity field on intermediate scales. It is likely that the linear model may have reached its maximum capacity of prediction.
}

 {The \borg{} derived non-linear model of the velocity field is physically more realistic; but the simpler model in \citet{Carrick:2015xza} has the virtue of having been compared to distance observations. We intend to do the same in a forthcoming publication for \borg{}. \borg{} is also self-consistently models redshift-space distortions, which brings additional information to disentangle galaxy bias from dark matter clustering. }

\begin{figure}
    \includegraphics[width=\hsize]{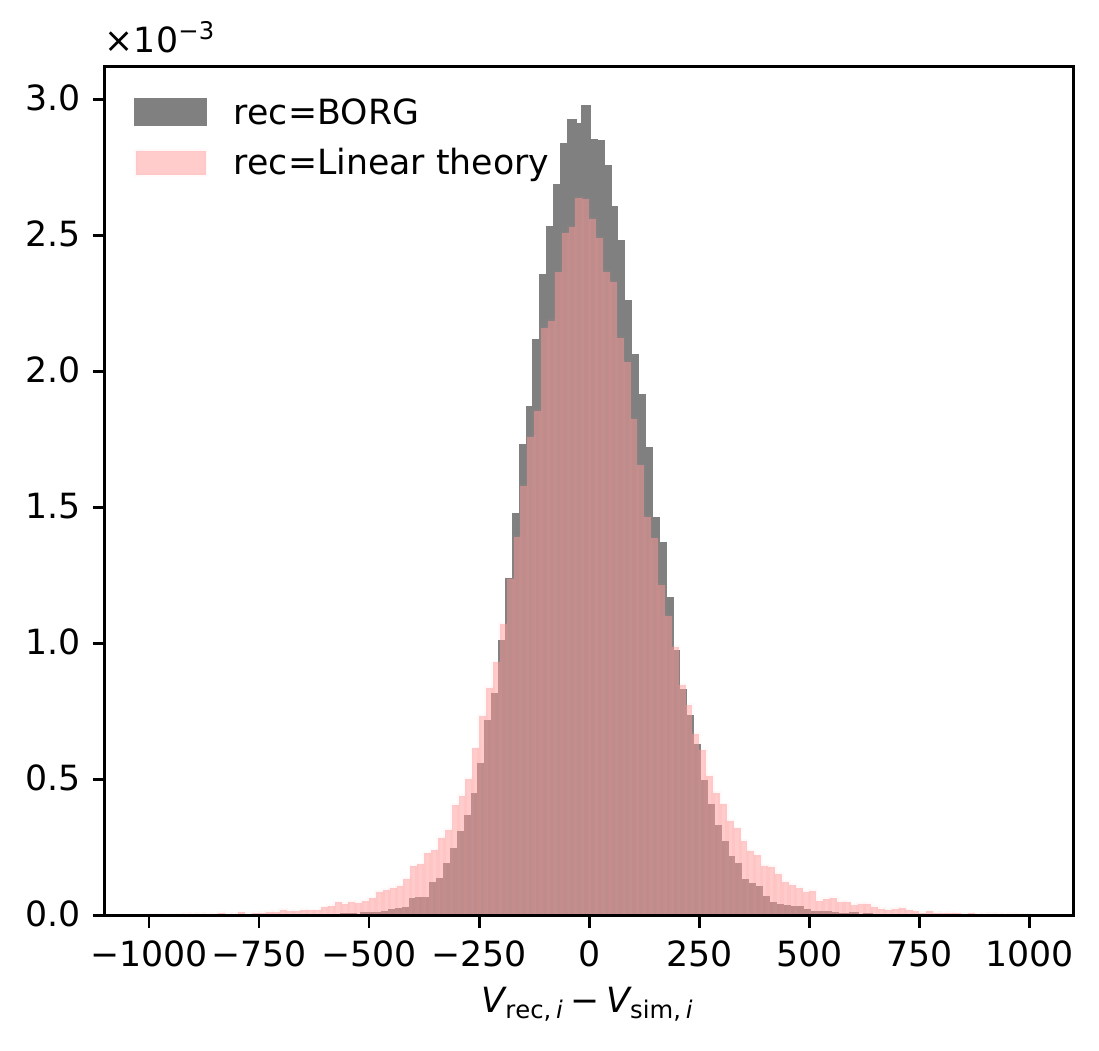}
    \caption{ {Tests of the BORG-PM algorithm on a $N$-body simulation: result for velocity field. We compute the velocity field using the Simplex-In-Cell algorithm using the entire set of particles of the $N$-body simulation and a resimulation from a single \borg{} sample. We compute the linear theory prediction for the velocity field based on the mock catalog as well. We show here the residual between each of the reconstructed velocity field and the one of the simulation. The difference is computed for each mesh element. We note that the residual distribution is narrower and have less heavy tails for the \borg{} reconstruction than for the linear theory one. In both cases, the velocity field is unbiased.}}
    \label{fig:borg_pm_velocity_residual}
\end{figure}

 {\subsubsection{Tests with an $N$-body simulation}
\label{sec:n_body_test}}

\begin{figure*}
    \centering
    \includegraphics[width=.8\hsize]{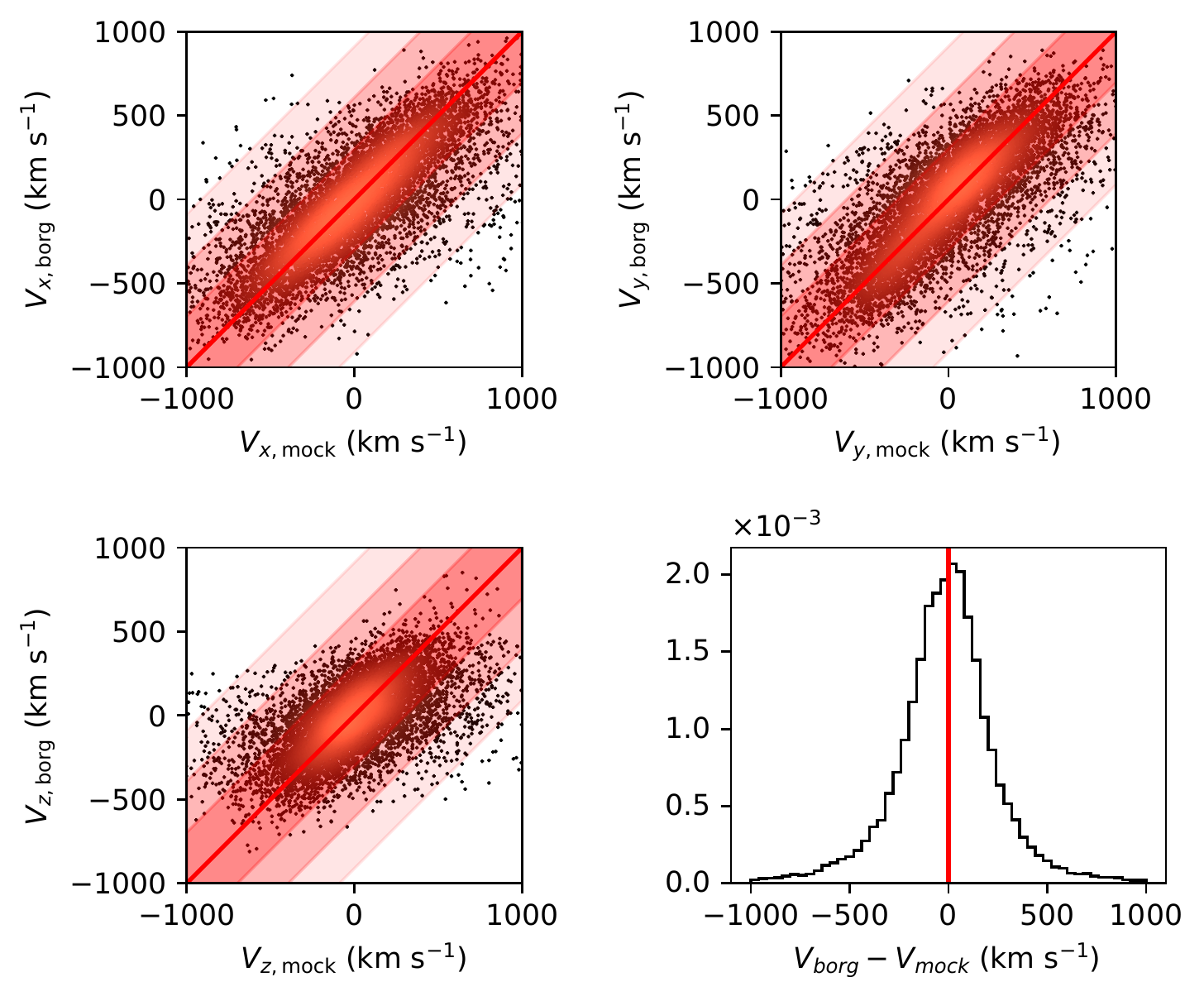}
    \caption{ {We show here the components ($x$ top-left, $y$ top-right, $z$ bottom left) of the velocities of the mock tracers as reconstructed from a single sample from \borg{} and compared to the ones provided by the original $N$-body simulation. The points are color coded according to their local density in the scatter plot to allow a better identification of the central part and the tails of the distribution. The diagonal red line is added for reference and to indicate unbiased velocity reconstruction. The lower right panel shows the histogram of the difference between the \borg{} reconstructed velocities and the original velocities of the mock tracers, for the three components.}}
    \label{fig:borg_pm_velocity_scatter_mock}
\end{figure*}

 {Though a full mock catalog analysis is beyond the scope of this article, we have run some tests on $N$-body simulations to showcase the correctness of the \borg{} reconstruction algorithm. Using Gadget-2 \citep{Springel2005}, we have generated an $N$-body simulation covering a cubic volume with a side length of 200\Mpch{}, and sampled with $256^3$ particles. We have used the same cosmology as the one that was used for the \borg{} 2M++ reconstruction in \citet{Jasche2019_PM}. To simplify the test case, we have generated a mock catalog of 8~457 objects directly from the particles of the simulation. The corresponding number density is $\sim 8-10$ times smaller than in the observational data. We have then run a BORG-PM reconstruction with 20 time-steps from $z=50$ to $z=0$ linearly distributed according to the scale factor, with a grid of $32^3$ elements to represent initial conditions. This corresponds to a spatial resolution of 6.25\Mpch{}, which is two to three times lower than the one used for the run on observations.}

 {%
We show in Figure~\ref{fig:borg_pm_density_reconstruction} a slice of the density field of the $N$-body simulation used for the test (top-left panel), as well as the result of the nearest-grid point assignment of the objects of the mock catalog in that same slice (top-right panel). We also show the density contrasts (bottom-left panel) computed from a single \borg{} sample, after the burn-in phase. We conclude with a difference plot highlighting the difference between the density field shown in the top-left panel and the bottom-left panel. We note the qualitative agreement between that sample and the density contrasts of the full simulation. In Figure~\ref{fig:correlation_sim_borg_density}, we give a quantitative comparison derived from the entire posterior distribution. The assessment is done by computing the cross-correlation between the \borg{} density field and the simulation density field and the variance between those two fields. They show that the \borg{} density field is unbiased and tightly correlated on large scales.}

 {%
Figure~\ref{fig:borg_pm_velocity_residual} illustrates the distribution of the differences between the velocity field of the simulation and the one provided by \borg{} inference. The comparison is done for a single sample obtained by \borg{} and for linear theory directly from the density contrast obtained from the mock catalog (linear theory). The velocity fields are computed using the Simplex-In-Cell algorithm at high resolution and downgraded at $32^3$ using local averages. 
In both cases, the reconstructions are unbiased but linear theory exhibits heavier tails for the residuals, leading to a typical width of 200~\kms{}. On the other hand, \borg{} has residuals with a typical width of 139~\kms{}. To further check the unbiased-ness, this time using the mock tracers themselves, we show in Figure~\ref{fig:borg_pm_velocity_scatter_mock} how the reconstructed velocities, which are obtained through the \borg{} velocity field, compares to the original mock tracer velocities directly taken from the $N$-body simulations. The first three panels (top-left, top-right, lower-left) give the scatter plots for each component of the velocities for each mock tracer. The lower-left panel shows the histogram of the residual velocities for the three components together.
}

 {We conclude from this test that the peculiar velocities reconstructed with the \borg{} algorithm are unbiased. They also have typically smaller errors compared to velocities derived using a linear perturbation theory approach.} 

\subsection{Implementation of BORG on 2M++}

\begin{figure*}
  \includegraphics[width=\hsize]{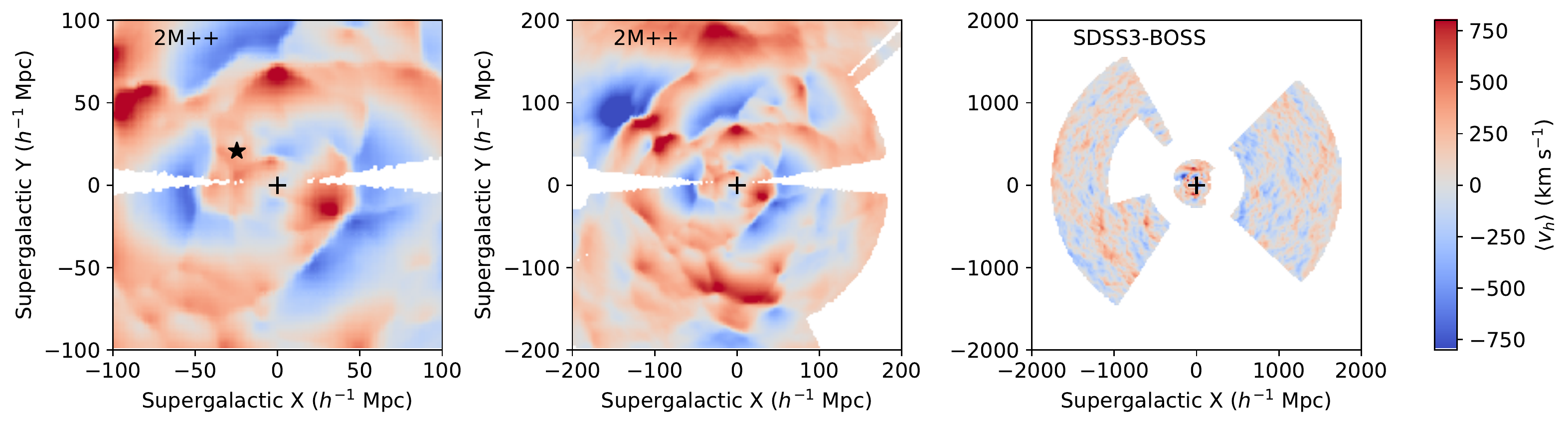}
  \caption{\label{fig:borg_vel}Global properties of the velocity field that is available from the \borg{} inference framework applied on the 2M++ compilation and on the SDSS-III/BOSS sample of galaxies. The black cross gives the position of the observer. Left panel: velocity field derived from \borg{} in the Supergalactic plane. The black star indicates  the spatial position of NGC 4993 which is the host of GW170817. Middle panel: Spatial distribution of the available model of the velocity field in the Supergalactic plane B$=0^\circ$ for the volume of 2M++. Right panel: Same as middle panel but for the SDSS-III/BOSS survey. The part at the center is the region from 2M++.}
\end{figure*}

\begin{figure*}
    \centering
    \includegraphics[width=.8\hsize]{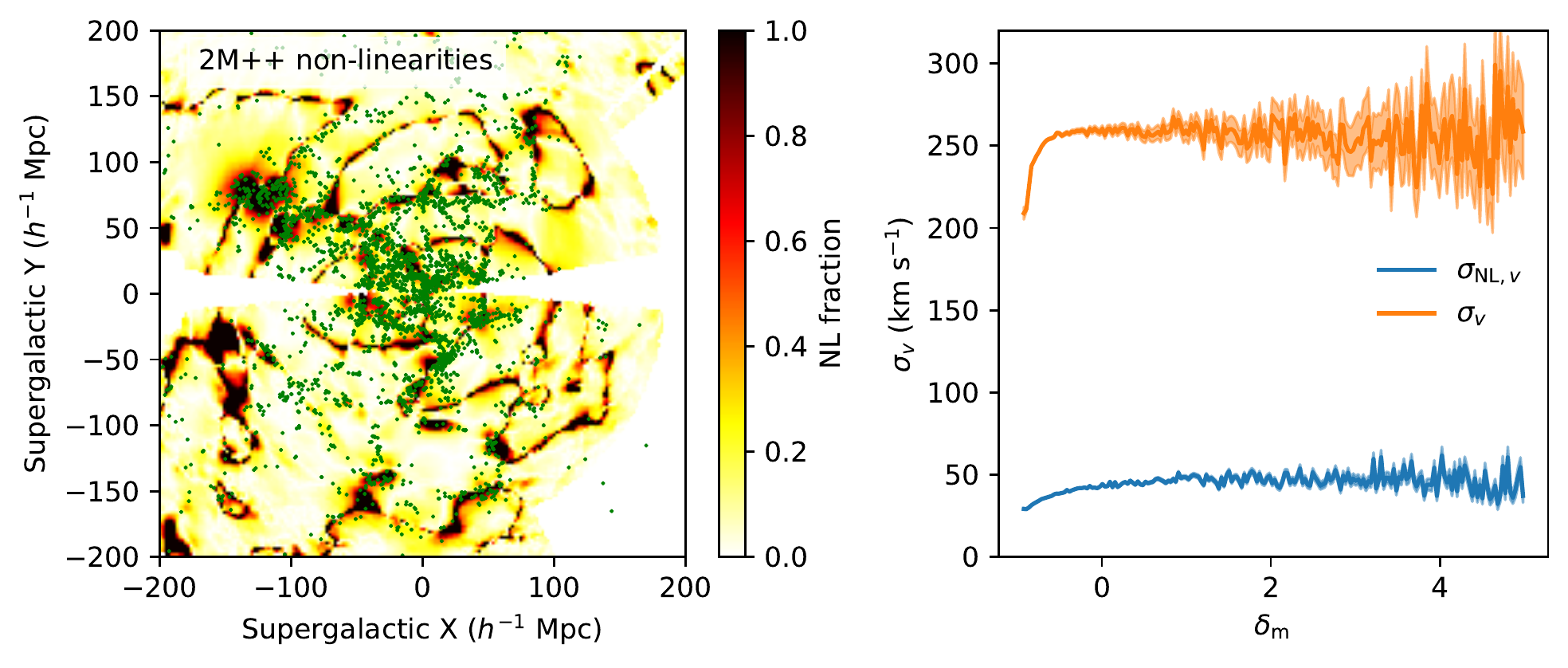}
    \caption{Non-linearities (NL) modelled by the \borg{} dynamical model. \label{fig:borg_nl} Left panel: the relative difference between the velocity field derived from the gravity field assuming linearity and the non-linear velocity field produced by \borg{} with respect to the total  {for the ensemble mean fields}. Right panel: standard deviation of the velocity field for each density bin (orange curve), and standard deviation of the non-linear velocity,  {$(v_{r,\text{BORG}}- v_{r,\text{linear}})$}, for each density bin (blue curve). We see here that the non-virial, non-linear component of the velocity field can represent a substantial portion of the cosmic velocity field: up to   {$\sim$25\%} of the standard deviation of the total velocity field when the local matter density {exceeds} $\delta_\text{m}\gtrsim 4$.  }\label{fig:borg_vel_nl}
\end{figure*}

 In Figure  \ref{fig:borg_vel} (left panel), we show the velocity field in supergalactic coordinates for 2M++ along with the starred spatial position of NGC 4993. The middle panel in Figure \ref{fig:borg_vel} indicates the estimate of the velocity field from 2M++ in the Supergalactic coordinate system. Additionally, we further show in the right panel the velocity field as inferred from the SDSS-III/BOSS survey. The latter uses a simpler dynamical model based on first order Lagrangian perturbation theory \citep{Lavaux2019_BORG}. The contribution of the non-linear velocity component captured by \borg{} is shown in Figure ~\ref{fig:borg_vel_nl}. This non-linear velocity component is the one captured by the particle mesh solver of \borg{}, although it is not yet sufficient to resolve virial motions in large scale structures. We denote this regime by "NL" for non-linear. It corresponds to the intermediate ``gray" regime when a linear modeling of the velocity field is not sufficient while still not describable without a full non-perturbative description of the non-linear dynamics. To assess its importance we consider two residuals. The left panel shows the relative contribution of the pure NL velocity from \borg{} with respect to the total velocity contribution  {by evaluating the expression $|(v_{r,\text{BORG}}-v_{r,\text{linear}})/v_{r,\text{BORG}}|$}.
 It is derived from the difference between the total contribution and the velocity field derived from the gradient of the gravitational potential of the matter density field.  {We note that the difference between the two fields may originate from at least two reasons: either the linear theory approximation underestimates the amplitude of flow in infalling regions, or it unphysically overestimates the velocity flow close to the peak of the matter density contrast. } The right panel shows the estimated standard deviation of the non-linear velocity field, both the total velocity and the residue obtained by taking the difference between the total non-linear and the linear component of the velocity field.  {This quantity corresponds to the second moment of the probability density function, conditioned on  $\delta_\text{m}$. We note that this moment does not correspond to the full marginal distribution, and thus it is possible that the conditional standard deviation is reduced compared to the marginal standard deviation.} We compute this standard deviation from each bin of matter density field. We note that the ratio of these two standard deviations of the total velocity field and the non-linear component component of the velocity field can go up to  {$\sim$ (20-30)}\% at $\delta_\text{m}\sim4$, indicating that the non-linear component is far from being negligible.

We point that the velocity and density field inference is largely independent of the Hubble constant. Our working coordinate system is using the arbitrary value of $H_{0}=100$~\kmsMpc{}. In this coordinate system, the value of $H_0$ is irrelevant for computing the dynamics and the observable in the observer coordinate frame. However, what matters is the redshift evolution of $E(z)$,  {which only depends on relative expansion history. For the portion of Universe that is considered here, the only relevant parameters are $\Omega_\text{m}=0.315$ and $\Omega_\Lambda=0.685$ \citep{Aghanim:2018eyx}.}
The only place where the absolute value of $H_0$ enters is in the a-priori power spectrum of primordial fluctuations that was used to run the analysis. This is however at least a second-order effect as this only governs the way \borg{} has to fill the missing information in the data space, but not the actual mass distribution that is inferred on large scales.

\section{Algorithm for the peculiar velocity correction for standard sirens}\label{paves}

The observed GW signal from a network of GW detectors is a probe of the distance to the source as discussed previously. The GW signal parameters comprise the source parameters (such as the chirp mass, individual masses of the compact objects, spin of the individual compact objects, the object's tidal deformability) and parameters that relate to the observer of the GW source such as the luminosity distance $d_L$, sky location $\hat n$, polarization angle $\psi$, and the inclination angle $i$. It is this latter set that is useful to estimate the value of $H_0$ \citep{Dalal:2006qt, 2010ApJ...725..496N,Abbott:2017xzu}.  The GW strain $h_+$ and $h_\times$ provides estimates of the luminosity distance and the inclination angle of the GW source. The sky location of the GW source is obtained using the arrival times of the GW signal at different detectors, as well as their antenna function \citep{2009NJPh...11l3006F, Nissanke:2011ax, Fairhurst:2010is, 2011CQGra..28l5023S, 2012PhRvD..85j4045V, Nissanke:2012dj}. The redshift of the GW source may be measured by identifying the host galaxy using the electromagnetic counterpart from the GW sources.
 
We propose to estimate the peculiar velocity of the host galaxy by estimating the large scale flow $v_h$ from  \borg{}, and the small scale motion as a stochastic velocity dispersion using the fitting form given in Eq.~\eqref{vvir}. We note that $v_h$ is itself composed of the velocity field at the linear order in perturbation, with an additional at least 30\% correction from non-linear evolution provided by \borg{} for density $\delta_\text{m}>0$. Using the location of the identified host galaxy (from the electromagnetic counterpart), we estimate the corresponding posterior Probability Density Function (\textit{pdf})  of the velocity field component $v_h$ from $118$ realizations of \borg{} for this galaxy. The virial velocity component is estimated using the mass of the galaxy halo in  Eq.~\eqref{vvir}. The \textit{pdf} of the non-linear velocity field $v_{vir}$ is assumed to be Gaussian distributed. The combined posterior of the peculiar velocity  $v_p= v_h + v_{vir}$ can be obtained by convolving the \borg{} posterior of $v_h$ with the \textit{pdf} of $v_{vir}$.

\begin{figure}
 \includegraphics[trim={0.2cm 0.2cm 0.2cm 0.2cm}, clip, width=1.\hsize]{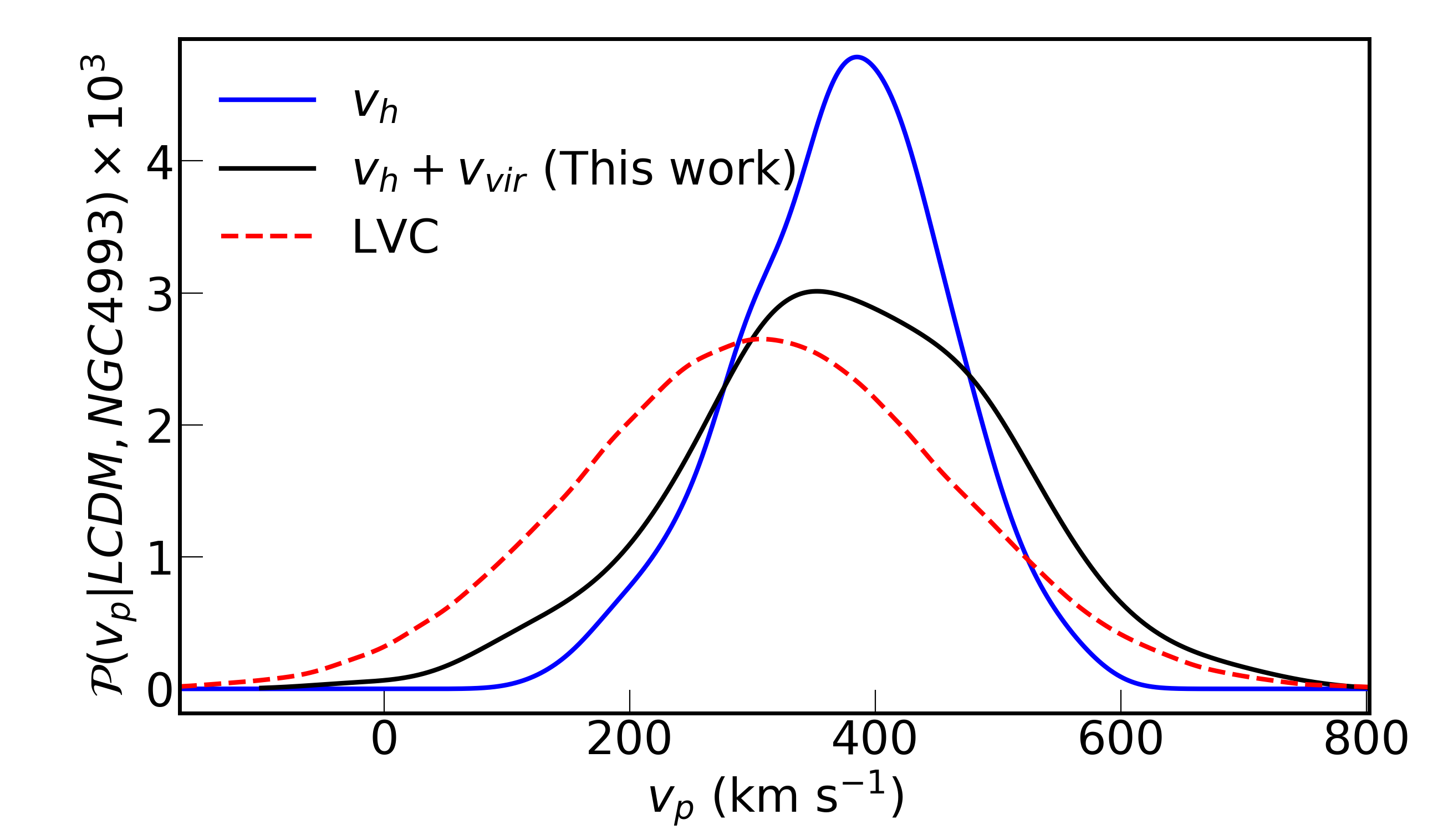}
\caption{The posterior of the peculiar velocity of NGC 4993. The blue curve displays the large scale flow, $v_p$, inferred from \borg{} while the black curve gives the required total velocity including the virial component within the halo. The posterior  of the velocity used by the LIGO Scientific Collaboration and Virgo Collaboration (LVC) is shown with red dashes.}
\label{fig:vel}
\end{figure}

By combining the measurements from GW data, $d_{GW}$, the redshift of the host galaxy, $\hat z$, and the peculiar velocity estimate, $v_p$, and luminosity distance $d_L$, we can obtain the posterior  of $H_0$ for $\mathbf{n}$ GW sources by using the Bayesian framework
 \begin{multline}\label{posterior_h0}
P(H_0|\, \{d_{GW}\},\, \{\hat z\}) \propto 
    \prod_{i=1}^\mathbf{n}\int \text{d}d^i_{L}\, \text{d}{v^i_p} \ \mathcal{L}(d^i_{L}| H_0, v^i_p, z_i, \vec{\hat{u}}_i, d^i_{GW})\ \times \\
    \mathcal{P}(z_i|\vec{\hat{u}}_i)\, \mathcal{P}(v^i_p| M,\vec{\hat{u}}_i)\ \Pi(H_0),
\end{multline}
where $\vec{\hat{u}}_i(\text{RA}_i,\text{Dec}_i)$ is the sky position,\footnote{RA and Dec respectively denote the right accession and declination.} $\mathcal{L}$ denotes the likelihood which is assumed to be Gaussian, $\mathcal{P}(z_i|\vec{\hat{u}}_i)$ is the posterior of the redshift estimate at the GW source $\hat {u}_i$, $\mathcal{P}(v^i_p| M, \vec{\hat{u}}_i)$ is the peculiar velocity estimate using the cosmological method, and $\Pi(H_0)$ is the prior on the value of $H_0$. 



\section{ {Applying our method to GW170817}}\label{results}

\subsection{Peculiar velocity estimate for NGC 4993}

NGC 4993 is the host galaxy of the event GW170817 (the merger of two neutron stars) which was discovered by the LIGO Scientific Collaboration and Virgo Collaboration (LVC) \citep{TheLIGOScientific:2017qsa, Abbott:2017xzu}. 
 {The distance to the GW170817 inferred from the GW observation is $d_L=43.8_{-6.9}^{+2.9}$ Mpc \citep{TheLIGOScientific:2017qsa, Abbott:2017xzu} and recession velocity of its host NGC 4993 with respect to the CMB rest frame is $3327 \pm 72$ \kms{} \citep{Abbott:2017xzu, Crook:2006sw}}
The velocity estimate for NGC 4993 has two components, $v_h$ and $v_\text{vir}$. We estimate the $v_h$ component from  \borg{}, and the non-linear part of the velocity is obtained using the form mentioned in Eq.~\eqref{vvir}. NGC 4993 has a halo of mass about $M_h \sim 10^{12}\, \Msun$ \citep{Pan:2017jem, Ebrova:2018gtz}. The corresponding estimate of $\sigma_\text{vir}$ is about 100~\kms{}. The excess velocity component is included as an additional velocity dispersion assuming a Gaussian \textit{pdf} of variance $\sigma^2_\text{vir}$ with a zero mean.

\begin{figure*}
     \includegraphics[trim={1.5cm 0.5cm 2.cm 0cm}, clip, width=1.\hsize]{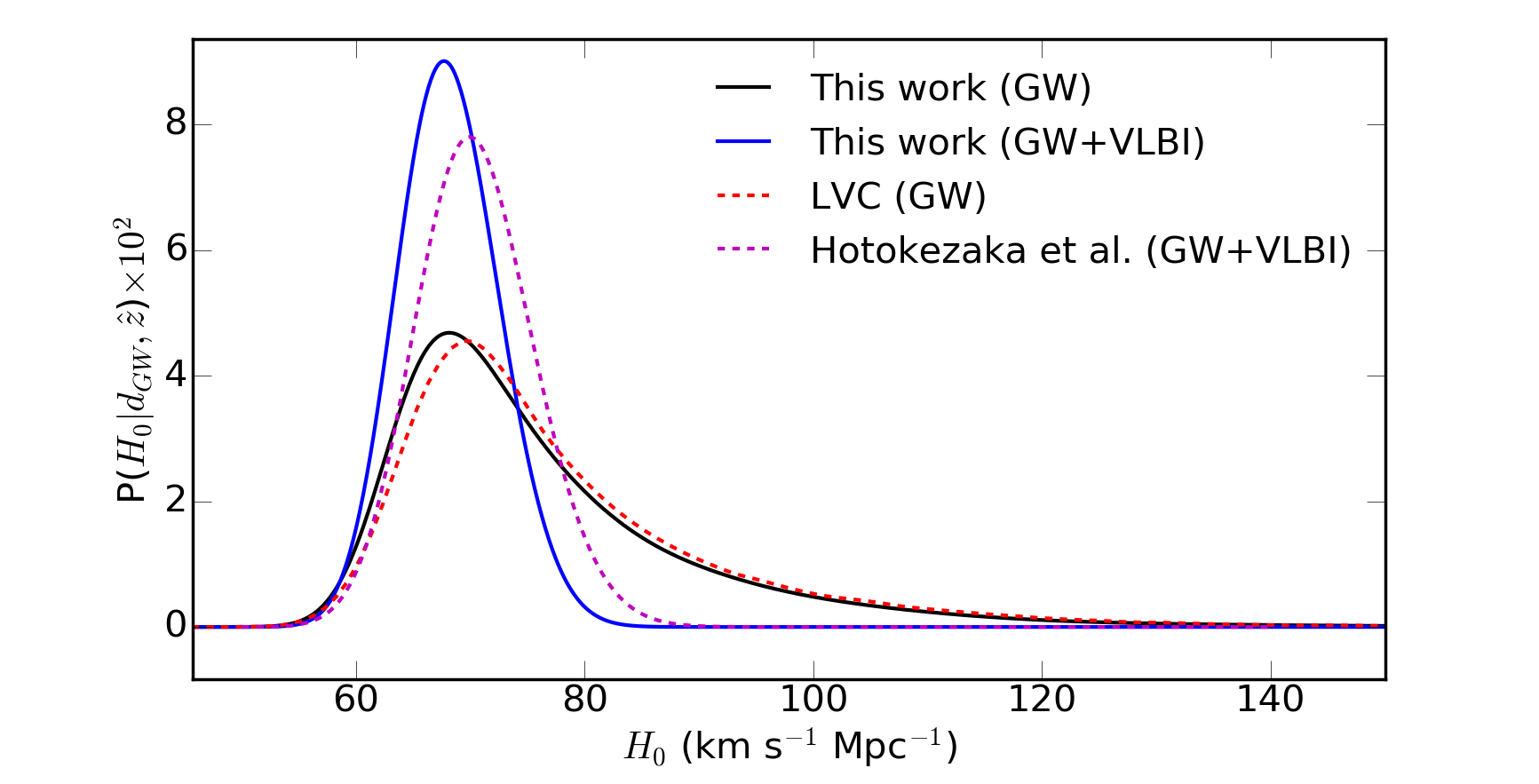}
\caption{ {The posterior  of $H_0$ for GW170817 is obtained using the peculiar velocity correction by our method and is shown in black when we use the GW data and in blue when using the GW and VLBI data. We also show the $H_0$ measurement from the LIGO Scientific Collaboration and Virgo Collaboration (LVC) in dashed red line \citep{Abbott:2017xzu} and the magenta dashes presents the result from \citet{Hotokezaka:2018dfi} from hydrodynamics simulation jet using GW+VLBI observation .}}
\label{fig:h0}
\end{figure*}

 {The posterior of the peculiar velocity from \borg{} is shown in blue color in Figure  \ref{fig:vel} for the Planck-2015 best-fit cosmological parameters \citep{Ade:2015xua}. The \textit{pdf} of the velocity field is non-Gaussian. The mean value of the velocity field is $\bar v_p= 373$~\kms{}, and the maximum a posterior value of the velocity field is $v^{MP}_p= 385$~\kms{}. The standard deviation of the velocity \textit{pdf} is $\sigma_h \sim 82$~\kms{}. The total standard deviation of the velocity  due to both $v_h$ and $v_{vir}$ is $\sigma_t= \sqrt{\sigma_h^2+ \sigma_{vir}^2} \sim 130$~\kms{}. The corresponding pdf for $v_p$ is shown by the curve in black color in Figure ~\ref{fig:vel}. The inferred value of the peculiar velocity of NGC 4993 differs from the value assumed by LVC \citep{Abbott:2017xzu}. \cite{Abbott:2017xzu} considered a velocity distribution which was Gaussian with a mean and standard deviation given by 310~\kms{} and 150~\kms{} respectively. The estimates of the velocity from our method predict a $24\%$ higher mean value of the velocity and about $13\%$ less standard deviation. The comparison of the distribution of the peculiar velocity between the LVC and our method is shown in red and black respectively in Figure ~\ref{fig:vel}.}

\subsection{Revised $H_0$ from GW170817} 
 {Using our new, more precise estimate of the peculiar velocity for the NGC 4993, we obtain the revised value of $H_0$ using the Bayesian framework mentioned in  Eq.~\eqref{posterior_h0} from the data of luminosity distance and inclination angle assuming either a high-spin or a low-spin prior on each compact object \citep{Abbott:2018wiz}. Figure  \ref{fig:h0} shows the corresponding posterior of $H_0$. The posterior of $H_0$ peaks at $H_0=68.3$~\kmsMpc{}, instead of $70_{-8}^{+12}$~\kmsMpc{} from the results of LVC \citep{Abbott:2017xzu} with the same $68.3\%$ credible interval as obtained by LVC. The marginal difference in the shape of the posterior between our method (black line) and LVC (red dashed line) arises only due to the difference in the velocity correction between our method and  LVC. In addition to calibration noise and peculiar velocity measurements, the main source of the error in the measurement of $H_0$ is due to the degeneracy between the inclination angle $i$ and the distance $d_L$  \citep{Abbott:2017xzu}. This acts as a limiting factor for the measurement of $H_0$ from GW170817, if there is no independent measurement of inclination angle. Critically, the addition of Virgo \citep{TheVirgo:2014hva} to the nearly co-aligned two LIGO detectors \citep{Abramovici325, Martynov:2016fzi} allowed for the improvement in the sky localization of the GW source and a possibility of improving the measurement of the inclination angle.}

A further improvement in the measurement of $H_0$ is possible if the uncertainty in the inclination angle $i$ can be reduced. A recent study has used the observed data of the electromagnetic counterpart to GW170817, the superluminal motion measured by the VLBI (Very Large Baseline Interferometry) observations \citep{Mooley:2018dlz} and the afterglow light curve data (e.g., \citet{2018ApJ...868L..11M}), to constrain the inclination angle of GW170817. Using the constraints on the inclination angle, \citet{Hotokezaka:2018dfi} obtained a revised value of $H_0$. We implement our velocity field correction method for the combined measurement of GW+VLBI, by using a flat prior on the value of inclination angle between $0.25$ rad $\leq i\, (\frac{d_L}{41 \text{Mpc}}) \leq 0.45$ rad. The revised $H_0$ by our method is shown in Figure  \ref{fig:h0} in blue color and is compared with the $H_0$ value from  \citet{Hotokezaka:2018dfi} (shown by the magenta dashed line). The error in the measurement of $H_0$ from GW+VLBI \citep{Hotokezaka:2018dfi} arises from several sources such as the GW data, the shape of the light curve, flux centroid motion, and the peculiar velocity. Our new estimate of the peculiar velocity improves the precision of the $H_0$ measurement with GW+VLBI from $70.3^{+5.3}_{-5.0}$ \kmsMpc{} to $68.3^{+ 4.6}_{-4.5}$ \kmsMpc{}.

\section{Conclusions and future prospects }\label{conclusion}

Cosmology with gravitational waves is a newly emerging field carrying enormous potential to explore new aspects of the Universe and fundamental physics \citep{Saltas:2014dha,Lombriser:2015sxa, Lombriser:2016yzn, Sakstein:2017xjx, PhysRevLett.119.251301, Nishizawa:2017nef, Belgacem:2017ihm, Pardo:2018ipy,  Abbott:2018lct,Belgacem:2018lbp, Belgacem:2019pkk, Mukherjee:2019wfw,Mukherjee:2019wcg}. One of such primary science goals is the measurement of the Hubble constant from the GW sources if the redshift to the source can be identified from the electromagnetic counterpart or the cross-correlation with the galaxy surveys \citep{PhysRevD.93.083511, Mukherjee:2018ebj}. With the current GW detectors such as Advanced LIGO \citep{Martynov:2016fzi}, and Advanced Virgo \citep{TheVirgo:2014hva}, recent forecasts predict the ability to measure $H_0$ with an accuracy of less than $2\%$  {with about $50$ BNS events} \citep{Chen:2017rfc, Feeney:2018mkj}. To achieve such a precise estimation of $H_0$, it is essential to accurately correct for the peculiar velocity of the hosts of GW sirens. This is particularly the case for very low redshift sources, which will have high signal-to-noise ratio and hence significantly contribute to the joint posteriors for
ensembles of events. In addition, peculiar velocity corrections will be important for events with  more favourable signal to noise in the Virgo or additional
other detectors, as well as neutron star-black hole mergers \citep{2010ApJ...725..496N, Vitale:2018wlg}.  The absence of such a correction will affect both the accuracy and precision of the measurement of the value of $H_0$. For a typical value of peculiar velocity of about $\sim 300$~\kms{} for a GW host at redshift $z=0.01$, the contribution from peculiar velocity is comparable to the term related to the Hubble flow. The contribution becomes less (or more) severe at higher (or lower) redshift. As a result, we need to estimate the value of the peculiar velocity with sufficient accuracy in order to avoid any systematic bias and additional variance in the measurement of $H_0$. Though averaging over a large GW samples can lead to an unbiased estimate of $H_0$, the absence of a peculiar velocity correction will increase the error budget on $H_0$ due to the additional variance from the peculiar velocity contribution.  In such a case, one needs a larger number of GW samples $N_{gw}$ to beat the variance as $N_{gw}^{-1/2}$ \citep{2010ApJ...725..496N, Nissanke:2013fka,Chen:2017rfc, Feeney:2018mkj, Mortlock:2018azx}. Incorporating an accurate correction for the peculiar velocity of the host of GW sources, we can achieve faster and more economically both accurate, and precise measurements of $H_0$.

In this paper, we use a statistical reconstruction method to correct the peculiar velocity of the host of the GW sources. The peculiar velocity for the host galaxy arises from the gravitational potential of the cosmic density field. We estimate the posterior of the peculiar velocity for both the linear and the non-linear component. The large scale velocity flow is estimated using the Bayesian framework called \borg{}. The stochastic velocity dispersion of the source is obtained using a numerical fitting-form given in Eq. \eqref{vvir}, by using the mass of the halo of the host. By combining the results from peculiar velocity estimate for each source, their redshifts, and the inferred luminosity distance from the GW data, we obtain a Bayesian inference of the value of $H_0$ according to the framework discussed around Eq. \eqref{posterior_h0}.

We implement our method on GW170817 \citep{TheLIGOScientific:2017qsa} which is in the host NGC 4993 \citep{Pan:2017jem}.  The corresponding posterior distribution with the results from LVC \citep{Abbott:2017xzu} and GW+VLBI \citep{Hotokezaka:2018dfi} are shown in Figure  \ref{fig:h0} by the solid line in black and blue color respectively. While our correction marginally reduces the maximum a posteriori value of $H_0$ to $68.3^{+4.6}_{-4.5}$ \kmsMpc{}  for GW+VLBI, it slightly disfavours very low values of $H_0\lesssim 60$ \kmsMpc{} compared to the results from LVC. As we were preparing this manuscript for submission, an analysis appeared  \citep{Howlett:2019mdh} that implemented an alternative velocity correction approach to the host of GW170817, and  recovered a value of $H_0= 64.8^{+7.3}_{-7.2}$ \kmsMpc{}. The mean value of $H_0$ differs from the values obtained in this work ($H_0= 68.3^{+ 4.6}_{-4.5}$ \kmsMpc{}) and also from the value obtained by \citet{Hotokezaka:2018dfi} ($H_0= 70.3^{+5.3}_{-5.0}$ \kmsMpc{}). However, the value by \citet{Howlett:2019mdh} is consistent within the error-bars of both the estimations. Also another recent work \citep{Nicolaou:2019cip} has discussed the possible impact of  peculiar velocity on the standard siren measurements, and has obtained a revised value of $H_0= 68.6_{-8.5}^{+14}$ km
~s$^{-1}$~Mpc$^{-1}$ for the event GW170817.  {Furthermore, a recent work by  \citet{Boruah:2020fhl} also discusses the impact of line of sight marginalization for models derived using linear theory applied to spectroscopic data, calibrated with supernovae data, and velocity field derived from the distance data obtained from Tully-Fisher or Fundamental Plane methods.}

The velocity correction is readily available for GW sources with electromagnetic counterparts (such as binary neutron stars, black hole neutron star) in the cosmic volumes covered by the 2M++ \citep{Lavaux2011_TMPP,Jasche2019_PM} and the SDSS-III/BOSS surveys \citep[][see Figure ~\ref{fig:borg_vel}]{2011AJ....142...72E,Dawson2013,Lavaux2019_BORG}. The 2M++ volume covers galactic latitudes $|b|>10^\circ$ up to redshift $z\sim 0.05$; the SDSS-III/BOSS survey spans  redshifts $z\sim 0.2-0.7$ for the sky areas ($0^\circ<\text{Dec}<60^\circ$ and $120^\circ<\text{RA}<240^\circ$) or ($0^\circ<\text{Dec}<30^\circ$ and $|\text{RA}|<30^\circ$). We expect that, within a year, our method will be available for the SDSS-IV/eBOSS survey \citep{2016AJ....151...44D}. In the long term, with the availability of the nearly full sky data sets jointly from the upcoming cosmology missions such as DESI \citep[Dark Energy Spectroscopic Instrument,][]{Aghamousa:2016zmz}, and LSST \citep[Large Synoptic Survey Telescope,][]{2009arXiv0912.0201L}, our algorithm will be available for most of the low redshift GW sources which will be observed by the currently planned network of ground-based GW detectors \citep{Schutz:2011tw, Aasi:2013wya}.

\begin{acknowledgements}
SM would like to thank Emanuele Berti, Rahul Biswas, Stephen Feeney, Daniel Mortlock, Daniel Sclonic and Joseph Silk for fruitful discussions. SM and SMN would like to thank Hiranya Peiris and Will Farr for useful comments on the draft of the paper.
The Flatiron Institute is supported by the Simons Foundation.  SM is supported by the Lagrange postdoctoral Fellowship at Institut d'Astrophysique de Paris. The work of SM and BDW are also supported by the Labex ILP (reference ANR-10-LABX-63) part of the Idex SUPER, and received financial state aid managed by the Agence Nationale de la Recherche, as part of the programme Investissements d'avenir under the reference ANR-11-IDEX-0004-02. BDW thanks the CCPP at New York University for its hospitality while this work was completed. 
This work was supported by the ANR BIG4 project, grant ANR-16-CE23-0002 of the French Agence Nationale de la Recherche. SM and SMN are grateful for support from NWO VIDI and Projectruimte Grants of the Innovational Research Incentives Scheme (Vernieuwingsimpuls) financed by the Netherlands Organization for Scientific Research (NWO). FL acknowledges funding from the Imperial College London Research Fellowship Scheme. KH is supported by Lyman Spitzer Jr. Fellowship at Department of Astrophysical Sciences, Princeton University.
This work was granted access to the HPC resources of CINES  (Centre  Informatique National de l'Enseignement Sup\'erieur) under the allocation A0020410153 made by GENCI and has made use of the Horizon cluster hosted by the Institut d'Astrophysique de Paris in which the cosmological simulations were post-processed. We thank Stéphane Rouberol for running smoothly this cluster for us. This work is done within the Aquila. Consortium\footnote{\url{https://www.aquila-consortium.org/}} We acknowledge the use of following packages in this analysis:  Astropy \citep{2013A&A...558A..33A, 2018AJ....156..123A}, emcee: MCMC Hammer \citep{2013PASP..125..306F}, IPython \citep{PER-GRA:2007}, Matplotlib \citep{Hunter:2007},  NumPy \citep{2011CSE....13b..22V}, and SciPy \citep{scipy}. The authors would like to thank the LIGO scientific collaboration and Virgo collaboration for providing the data for the event GW170817.
\end{acknowledgements}

%
%

 \bibliographystyle{aa} 
 \bibliography{main} 
\end{document}